\documentclass[sigconf]{acmart}
\setcopyright{none}
\AtBeginDocument{%
  }

\copyrightyear{2025}
\acmYear{2025}
\setcopyright{cc}
\setcctype{by}
\acmConference[Websci '25]{Proceedings of the 17th ACM Web Science Conference 2025}{May 20--24, 2025}{New Brunswick, NJ, USA}
\acmBooktitle{Proceedings of the 17th ACM Web Science Conference 2025 (Websci '25), May 20--24, 2025, New Brunswick, NJ, USA}
\acmDOI{10.1145/3717867.3717918}
\acmISBN{979-8-4007-1483-2/2025/05}

\begin{document}

% \title{DALiSM: A Discourse Analysis Framework for Legislative and Social Media Texts}
\title{A Discourse Analysis Framework for Legislative and Social Media Debates}
\author{Arman Irani}
% \email{airan002@ucr.edu}
\affiliation{
\institution{University of California, Riverside}
\city{Riverside}
\state{California}
\country{USA}}

\author{Ju Yeon Park}
\affiliation{
\institution{The Ohio State University}
\city{Columbus}
\state{Ohio}
\country{USA}}

\author{Michalis Faloutsos}
\affiliation{\institution{University of California, Riverside}
\city{Riverside}
\state{California}
\country{USA}}

\author{Kevin Esterling}
\affiliation{\institution{University of California, Riverside}
\city{Riverside}
\state{California}
\country{USA}}


\begin{abstract}
 How can we capture the dynamics of deliberation in a debate? In an increasingly divided and misinformed world, understanding the relationship between who is arguing and what they are arguing about is becoming critical for fostering a meaningful exchange of ideas. Given the vast array of available platforms for people to express their viewpoints and deliberate on issues, how can we develop methods to accurately analyze these processes? Luckily, there is an abundance of debate data available, ranging from: (a) formal proceedings, such as committee hearings in legislatures, to (b) online discussion forums, such as Reddit. Here we introduce DALiSM, a data-driven argument-centric framework, to analyze discourse dynamics in diverse and multi-party spaces at scale. We develop methods to harness and extend the state-of-the-art in computational argumentation for: (a) identifying arguments from long-form raw texts, (b) calculating the intensity of deliberation,  and (c) modeling the evolution of discourse over time. We deploy our framework as a comprehensive and interactive dashboard for dynamically viewing the outputs of DALiSM to clearly understand the nature of a discourse event. To showcase the importance and utility of DALiSM, we apply our framework to U.S. congressional committee hearings from 2005 to 2023 (109th to 117th Congresses), and to selected Reddit communities from 2008 to 2023. This case study reveals substantive insights into deliberative behavior in \textit{online} and \textit{offline} spaces.
\end{abstract}

%    <concept>
%        <concept_id>10010405.10010455.10010461</concept_id>
%        <concept_desc>Applied computing~Sociology</concept_desc>
%        <concept_significance>500</concept_significance>
%        </concept>
%    <concept>
%        <concept_id>10002944.10011123.10011673</concept_id>
%        <concept_desc>General and reference~Design</concept_desc>
%        <concept_significance>300</concept_significance>
%        </concept>
%    <concept>
%        <concept_id>10003120.10003145.10003151.10011771</concept_id>
%        <concept_desc>Human-centered computing~Visualization toolkits</concept_desc>
%        <concept_significance>100</concept_significance>
%        </concept>
%  </ccs2012>

% \ccsdesc[500]{Applied computing~Sociology}
% \ccsdesc[300]{General and reference~Design}
% \ccsdesc[100]{Human-centered computing~Visualization toolkits}




\maketitle

\begin{figure}[htbp] 
\centering
\includegraphics[width=\linewidth]{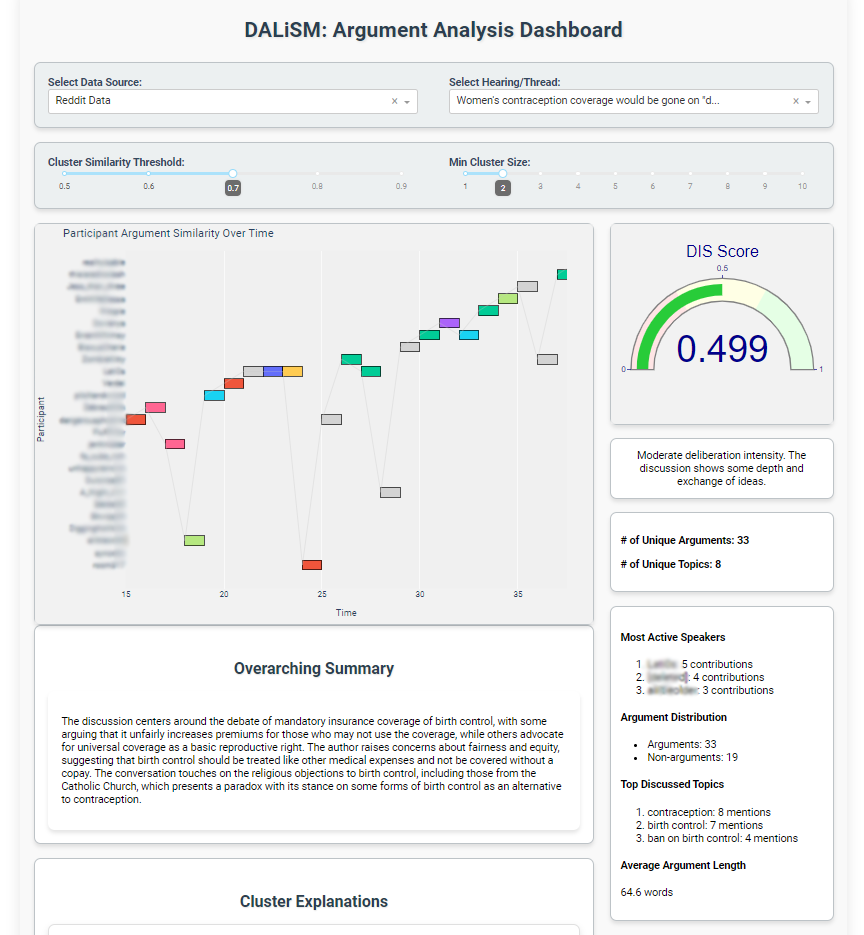}
\Description{A dashboard interface from the DALiSM platform showing deliberation dynamics during a discourse event. The y-axis represents participants, while the x-axis represents time. Colored boxes correspond to arguments, grouped by clusters based on argument similarity. The interface includes features such as automatic summaries, argument clustering, and deliberation quality metrics.}

\caption{DALiSM platform visualizing Reddit discussions: Participants on the y-axis, time on the x-axis. Each colored box shows an argument, with matching colors representing similar points. The system summarizes content and measures discussion quality. Usernames are blurred for privacy.}
\label{fig:delib}
\end{figure}

\section{Introduction}

Understanding the dynamics of discourse and how environments shape these discussions has never been more important as polarization, toxic rhetoric, and misinformation become increasingly prevalent. How we exchange and process ideas depends heavily on the environment in which the discourse takes place \cite{jones2015discourse}. Deliberation in a Reddit thread, referred to in this paper as \textit{online} discourse is often informal, asynchronous, and largely anonymous, contrasting sharply with deliberation taking place in legislative proceedings, which falls under the scope of what this paper terms \textit{offline} discourse which is often public-facing, structured, synchronous, and formalized. Factors influencing these processes are vital for downstream analyses of how digital and institutional spaces foster or degrade the democratic exchange of ideas and the nature of what is being expressed.

\textbf{Problem:} 
How can we model the argumentation dynamics of discourse? This is the question that drives our research. Deliberative events, while complex and multidimensional, contain a treasure trove of information that aids researchers and policymakers in better understanding deliberative democracies. A crucial element that is often overlooked in this current landscape of political methodology is the argumentation dynamics of such processes. Scholars typically use topic models and stance detection models to analyze the content and its transformation throughout a dialogue. We argue that, although these methods capture informative and meaningful analysis, a core element missing is an \textit{argument-centric} approach to these topic and stance models. Furthermore, the challenge of being able to process, understand and visualize argumentation dynamics at scale is necessary as millions of deliberative events take place every year. In this paper, we focus on an extremely important deliberative process, U.S. congressional committee hearings, which are central to policymaking in the U.S. Congress. Although these proceedings are critical for the function of Congress, there has been little progress in using computational text analysis methods to analyze them for their deliberative quality \cite{big_data}. Thousands of hearings take place every year in the United States, and it is impossible to make sense of such a large amount of information in its raw form. We demonstrate our approach's ability to capture deliberation in various environments by showcasing our approach applied to both these committee hearings and Reddit discussions. Here are some motivating questions:

a) What are the arguments in the debate by each participant?

b) How do the arguments evolve during the deliberation?

c) Does the discussion remain focused on its topic?

d) What is the overlap in arguments among debaters?

\textbf{Previous Work:} Despite the abundance of research analyzing social media discourse, there has been little to no work done to provide a framework of comparison across \textit{online} and \textit{offline} platform debate. Traditionally, research has focused on the content of the debate, specifically on the quality and validity \cite{steenbergen2003measuring, big_data}, but not on the dynamics or factors affecting the deliberation. Furthermore, the current landscape of discourse analysis focuses more on straightforward topic modeling, or sentiment detection approaches which, although informative, usually fail to capture the complete picture. The abundance of data and the focus on tools for short-form text have created a gap in effectively analyzing deliberative behavior at scale. 
We review prior efforts in the Related Works section.

\textbf{Contribution:}
We propose DALiSM, a \textbf{D}iscourse \textbf{A}nalysis framework for \textbf{L}eg\textbf{I}slative and \textbf{S}ocial \textbf{M}edia debate. It is 
%a systematic and comprehensive
a systematic, comprehensive, and argument-centric
framework to quantify the dynamics of deliberation across different platforms. 
In essence, DALiSM processes unstructured debates and outputs meaningful metrics, insights, and visualizations for each deliberative event. The key unit of analysis in our approach is the ``argument'', defined semantically as a claim that is supported by at least one premise.
The novelty of the framework lies in the confluence of the three following characteristics:
(a) its design principles are rooted in the theory of deliberative democracy \cite{Habermas1984}, focusing on the role of argumentation and reason-giving in democratic politics; 
(b) it follows an argument-centric approach, using the argument as its foundational analytic unit; and
(c) it captures the interplay of arguments, topics, and participants holistically by introducing six novel metrics. As an additional contribution, we develop DALiSM as a publicly accessible web platform\footnote{Platform is accessible at: \url{www.dalism.dev}} as shown in Figure~\ref{fig:delib}.

The  platform provides a user-friendly wrapper around our methods in order to make them readily available as capabilities to researchers and practitioners. 
The key features of our platform include the following:
(a) it can handle both congressional hearings\footnote{Analyzing congressional hearings is relatively neglected in computer science but is of foundational interest to political science.} and online discussion forums data;
(b) it provides concise and insightful summaries and visualizations;
(c) it is interactive;
and (d) its user-friendly design makes it easy to use even by non-technical people.
Building on these capabilities, our ambition is to make the platform available to the research community and establish it as a reference point for analyzing deliberative discourse. By leveraging DALiSM's metrics and visualizations, we can deeply explore expressed arguments, assess their diversity, and how they flow and interact within debates. These methods facilitate the transformation of raw argumentation data into insightful and interactive outputs. To demonstrate the framework's effectiveness, we evaluate its performance using real-world data. Specifically we apply DALiSM to debates in U.S. congressional committee hearings and discussions on Reddit, focusing on two controversial topics: {\it abortion} and {\it genetically-modified organisms (GMOs)}.

Our analysis reveals distinct evolutionary patterns in deliberation across platforms. Congressional hearings exhibit a steady increase in argument diversity, reflected by positive discourse evolution trends. In contrast, Reddit discussions show a slightly regressive trajectory, indicating a lack of sustained semantic progression over time.  Additionally, we find that Reddit discussions are significantly more volatile than congressional hearings with volatility values up to six times higher. While congressional hearings maintain a stable and linear progression, Reddit discussions fluctuate considerably, particularly during the final phases of discourse. Further we find that congressional hearings foster more structured but less participatory deliberation, while Reddit supports broader participation and more frequent argumentation.

\section{Methodology}

We present algorithmic solutions to substantiate our argument-centric framework. Specifically, we introduce: (a) a novel sliding window technique to overcome challenges in identifying nested arguments within long texts, (b) methods for grouping similar arguments, (c) a novel set of six metrics for quantifying deliberation,
(d) a method for quantifying the evolution of arguments in a discourse event, and finally, (e) an interactive dashboard for viewing these deliberation dynamics, as an engineering contribution.

\subsection{Argument Detection}
In this work, we leverage the capabilities of our open-sourced tool WIBA \cite{wiba}, which comprehensively identifies the presence of an argument, the topic being argued, and the stance of that argument. Each of these three components is referenced as WIBA-Detect, WIBA-Extract, and WIBA-Stance throughout this work. These three fine-tuned LLama 3 8B large language models, which we developed, were demonstrated to surpass current state-of-the-art monological argument detection methods with an $F_1$ score of 80-86\% for argument detection and 71-78\% for stance detection across various benchmarks \cite{wiba}. Furthermore, we performed an extensive evaluation of the tool's accuracy on informally structured arguments and formally structured arguments, with a detection accuracy of 76\% and 89\% respectively. For WIBA-Extract, the task of extracting the core topic being argued, we evaluated this component and demonstrated its ability to significantly outperform current na\"ive topic modeling methods, with an average cosine similarity score of an extracted topic to the gold standard of 74.2\%. This assessment was critical, as it allows us to build upon and extend our open-sourced computational argumentation tools in a systematic and robust manner.

\begin{figure}[htbp] 
\centering
\begin{minipage}{\linewidth}
    \includegraphics[width=\linewidth]{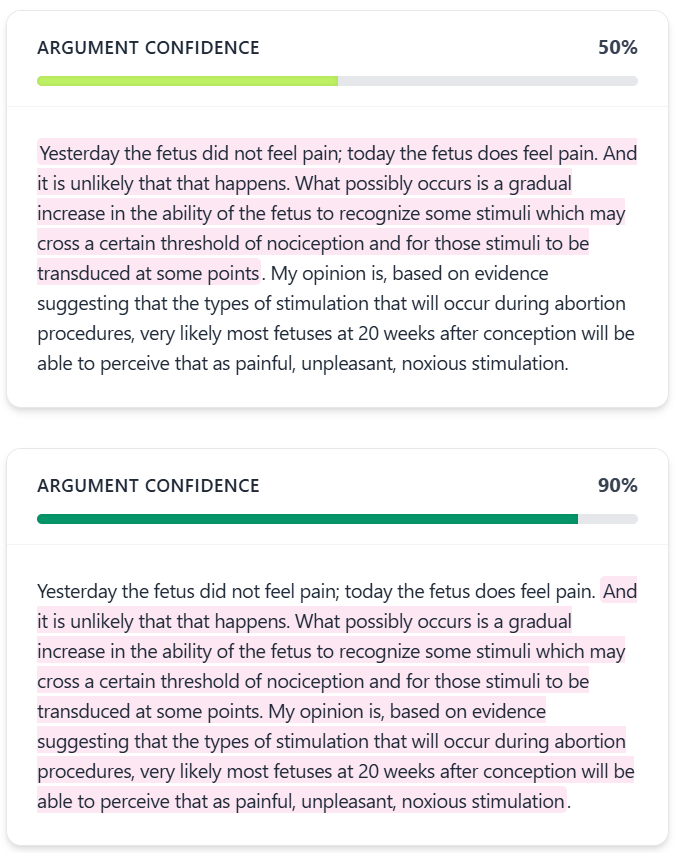}
    \Description{A visualization of the automated sliding window Argument Detection process. Two text windows are shown, with the second window having a higher argument confidence score, indicating that it is more likely to be an argument compared to the first window.}
    \caption{This figure shows an example of our automated sliding window Argument Detection process. In this example, the second window has a higher argument confidence. Therefore we assume this text unit is more of an argument than the first window.}
    \label{fig:window}
    \centering
\end{minipage}
\end{figure}

\subsection{Argument Extraction within Long Texts}
To address the challenge of identifying arguments in long text sections where the exact span of an argument is unknown, we employ a \textit{sliding window} technique. For each statement or utterance made by a participant, we consider every possible $k$ consecutive sentences and call it a text unit to be analyzed. For this study, a window size $k$ of three was chosen, since the argument detection model WIBA-Detect \cite{wiba} was trained on a dataset with an average text unit size of three sentences. We define the step size to be one, so the text unit will move forward by one sentence.

The argument detection model, WIBA-Detect, analyzes each text unit, determines whether those three sentences form an argument, and assigns a binary label to the text unit: \{NoArgument, Argument\}. Along with the label, it computes a confidence value for the decision, and those receiving a confidence level above 0.5 are considered an argument. With a step size of one, there will be overlapping windows with varying confidence scores, as shown in Figure \ref{fig:window}. If two overlapping windows are both labeled as an argument, the window with the higher confidence retains its ``Argument'' label initially assigned, but the label for the other window is adjusted to be ``NoArgument.'' This ensures that only the most obvious argumentative text unit is labeled as an argument, but not the text surrounding it.

\subsection{Thematic Argument Similarity}
\label{section:arg_theme}
A key to understanding the substantive nature of discourse is the ability to measure the degree of similarity between arguments. To do so, we use the state-of-the-art Sentence Transformer model, `all-mpnet-v2', which has been fine-tuned on 1 billion sentence pairs using contrastive training \cite{reimers-2019-sentence-bert}. This model calculates the semantic similarity between two sentences or paragraphs (up to 384 words) on a scale of 0 to 1. A score of 0 means the two texts are completely dissimilar, and a score of 1 means the two texts are identical.

When testing on the Best-Worst Scaling (BWS) Argument Similarity corpus provided by Ubiquitous Knowledge Processing Lab (UKP), we obtain a Cosine $F_1$ score of 70.2\%. This result indicates a strong ability to determine the narrative similarities between arguments, as the BWS corpus consists of 3,400 pro/con stance arguments across eight controversial topics \cite{bws}.

\subsection{Quantifying Deliberation Intensity}

How diverse are the arguments or ideas that are exchanged in a discourse event? We define  \textit{Deliberation Intensity} as a metric to quantify the amount of deliberation occurring at both participant and discourse event levels. We are not suggesting that the Deliberation Intensity is a proxy for deliberation quality, such as what is measured in the Discourse Quality Index \cite{steenbergen2003measuring}, but rather it captures the variety of arguments made in a discourse \cite{ArguSense}.

We propose to measure the Deliberation Intensity of an event, which analyzes the structural and participatory properties of debates to create an informative profile of the event. The structural properties of deliberation encompass features related to the composition of the debate, such as narrative clusters, outliers, and inter-cluster distances. Participatory properties reflect the number of participants and arguments contributing to the debate. Our proposed metric produces a comprehensive score ranging from 0 to 1, offering an interpretable measure for quickly assessing debate intensity. While this method proposes a systematic approach for calculating the intensity of deliberation, we recognize the inherent challenges and complexity of this task and acknowledge the possibility for alternative frameworks offering complementary insights.  
Although one could consider different metrics,
we argue that our metric is a great baseline for describing deliberation as we observe in our cross-platform case study in Section \ref{sec:crossplatform}.

\subsubsection{Structural Properties} 
For our structural properties we utilize the clustering algorithm HDBSCAN, which generates non-uniform clusters, given our Sentence Transformer embeddings. We use these clusters to derive the features that make up the structural analysis of cross-platform debates. Since our embeddings capture semantic similarities of arguments with high precision, we extend an assumption that a cluster generated from a cohesive collection of semantically similar embeddings represents a distinct viewpoint or narrative. We propose two structural features that answer the following question: How \textit{diverse} and \textit{distinct} are the narratives in the debate?

\begin{figure}[htbp] 
\begin{minipage}{\linewidth}
    \includegraphics[width=\linewidth]{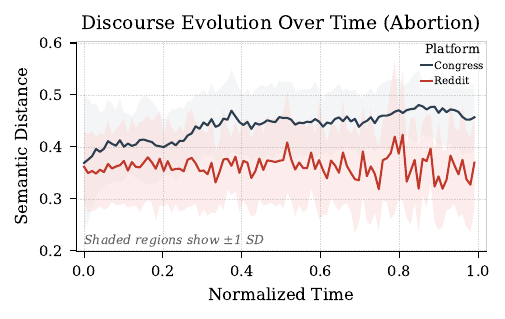}
    \Description{A line chart comparing the evolution of discourse on Reddit and congressional hearings regarding the topic of abortion. The chart shows higher volatility in Reddit discussions, especially toward the later stages, while congressional hearings follow a more stable trend with greater semantic differences over time.}
    \caption{The evolution of discourse on Reddit and congressional hearings regarding the topic of abortion. We observe more volatility in Reddit discussions, especially in the later stages, with an overall smaller semantic difference than the trend observed in congressional hearings.}
    \label{fig:evolution}
\end{minipage}
\end{figure}

\textbf{Narrative Diversity.} This feature serves as a metric for evaluating how \textit{diverse} the spectrum of arguments present in a debate is. Intuitively, this quantifies the number of unique narratives can be found in the discussion. The metric calculates the number of clusters identified by HDBSCAN.

To account for the expected non-linear growth of narratives in congressional hearings, we employ square root scaling, a technique motivated by empirical findings in online social media word fluctuations \cite{sano2015fluctuation}. It was observed that the growth of word frequencies in social media follows a Poisson-like distribution, characterized by a mean proportional to the square root of time. By applying square root scaling to our social media and congressional hearing analysis, we aim to stabilize the variance observed in the narrative dynamics and mitigate the impact of extreme values. While acknowledging potential limitations and considering alternative approaches, such as log-log or power-law scaling, square root scaling proves effective in this context, grounded in empirical evidence from related domains.

Additionally, this feature takes into consideration the amount of \textit{noise} present in the space. Noise is defined by the number of arguments not organized into any clusters\footnote{This is denoted by the cluster label -1 in the HDBSCAN algorithm.}, and the inverse of this \textit{noise} can be intuitively understood as a measure of narrative coherence. In a situation with low coherence, there are many outlier arguments that do not form clusters; in high coherence, there are few outliers as most arguments are well organized into clusters. The formulation for Narrative Diversity is: 

\begin{equation}
    \frac{\textnormal{\# Clusters}}{\sqrt{\textnormal{\# Arguments}}} * \textnormal{Coherence}
\end{equation}
And the formula for Coherence is:
\begin{equation}
    1 - \frac{\textnormal{\# Outliers}}{\textnormal{\# Arguments}}
\end{equation}

\textbf{Narrative Distinctness.} This feature serves as a metric for evaluating the degree of \textit{differentiation} among the identified narratives. This feature evaluates the pairwise distances between cluster centroid nodes, which represent distinct narratives. By combining both the mean and minimum distances and applying a square root transformation, the metric balances the global separation across clusters and also preserves local distinctiveness. The mean distance captures the overall dispersion of clusters across the narrative space, which provides a broad view of how they are distributed. The minimum distance highlights the closest proximity between two clusters, ensuring that even the most similar narratives have some degree of separation. A higher Narrative Distinctness value signifies greater separation between argument clusters (narratives), reflecting a higher degree of differentiation. Conversely, a lower value indicates reduced separation or greater homogeneity among narratives. The formula for calculating Narrative Distinctness is as follows: 

\begin{equation}
    \sqrt{(\textnormal{mean distance} * \textnormal{min distance})}
\end{equation}

\subsubsection{Participation Properties.} In addition to the structure, we incorporate participant features to account for both the density of debaters and the proportion of argumentative content present in the debate. This perspective acknowledges that a debate's intensity depends not only on its structural composition but also on the extent and diversity of participant engagement. 

\textbf{Debater Diversity.} This features serves as a metric to quantify the \textit{density} of speakers in argumentative contexts. This feature normalizes the number of unique debaters who present arguments using the square root of the total number of arguments. This accounts for diminishing returns, as we do not expect the number of unique debaters to grow proportionally with the number of new arguments presented. For example, in a discussion with 100 arguments, it is unlikely that there will be 100 different debaters. This normalization keeps the metric stable across discussions of varying sizes, ensuring it accurately reflects the diversity of participation relative to the number of arguments in a debate. The formula to calculate Debater Diversity is: 
\begin{equation}
   \frac{\textnormal{\# Debaters}}{\sqrt{\textnormal{\# Arguments}}}
\end{equation}

\textbf{Argumentativeness.} This feature serves as a metric to evaluate the \textit{ratio} of arguments in a debate. A higher value represents a discussion more focused on argumentation, while a lower value indicates a discussion lacking in-depth deliberation. We do not apply square root normalization in order to preserve the direct proportional relationship between the number of arguments and total number of statements. By keeping the denominator linear, the metric intuitively and accurately reflects the degree of argument concentration within a debate. The formula for Argumentativeness is:
\begin{equation}
    \frac{\textnormal{\# Arguments}}{\textnormal{\# Statements}}
\end{equation}

\textbf{Overall Deliberation Intensity Score.} Using the derived overall component scores, we combine the structure and the participant scores additively. Then we take the average of the structure and participation scores, respectively, as follows:

\begin{equation}
    \textnormal{Structure} = \frac{(\textnormal{Narrative Diversity + Narrative Distinctness})}{2} 
\end{equation}

\begin{equation}
    \textnormal{Participation} = \frac{(\textnormal{Debater Diversity + Argumentativeness})}{2}
\end{equation}

We finally calculate the Overall Deliberation Intensity Score as the weighted sum of the two fundamental properties. Each of the individual Structural and Participant properties have associated weights, $\alpha$ and $\beta$ which can be tuned depending on the researcher's preference for relative importance of each feature on the overall score. By default, these weights are set to 0.5 each, indicating equal importance: 
\begin{equation}
    DIS = \alpha * \textnormal{Structure} + \beta * \textnormal{Participation}
\end{equation}

\begin{figure*}[htbp] 
\centering
\begin{minipage}{\linewidth}
    \includegraphics[width=\linewidth]{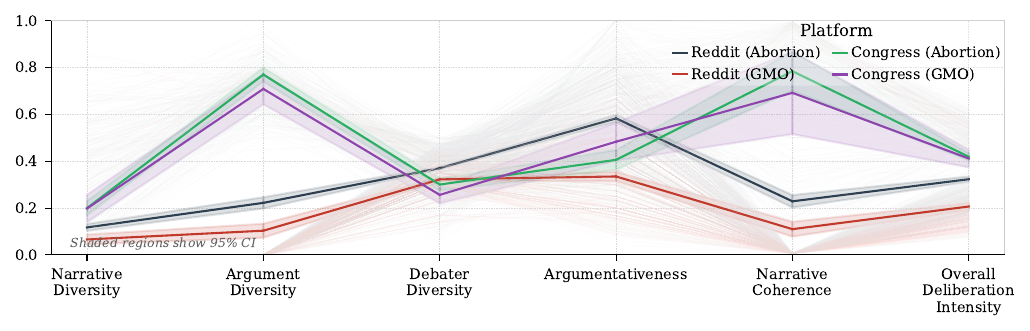}
        \Description{A line chart comparing average values and variances of deliberation metrics across U.S. Congressional hearings and Reddit threads. Congressional hearings show higher Argument Diversity and Narrative Coherence, indicating more structured and coherent deliberation compared to the less formal and more volatile discussions on Reddit.}

    \caption{ Establishing an informative deliberation profile with our metrics: We show the average and the variance of our metrics for each deliberation. Congressional hearings exhibit much higher Argument Diversity and Narrative Coherence than Reddit threads, which aligns with the nature of the two venues of deliberation. %We can observe the relative difference between the platforms and the impact that each Deliberation component has on the Deliberation Intensity Score.
    }
    \label{fig:ccdf}
    
    \centering
    
\end{minipage}
\end{figure*}

The Overall Deliberation Intensity Score is useful to analyze the nature of democratic deliberation. A higher score would suggest that more diverse views were densely exchanged by various participants in a deliberation session. This concept of intensity helps evaluating various deliberative activities in institutional decision-making processes or any public discussion for multiple reasons. In general, a group reflecting more diverse perspectives can make better decisions than a homogeneous group \cite{page2007}. In legislative deliberation, for example, the incorporation of diverse views can carry greater legitimacy for the lawmaking process \cite{schwindt2005}. Thus, this measure can enhance the systematic analysis of democratic deliberation in many directions. We visualize the relative impact of each feature on the Overall Deliberation Intensity Score in Figure \ref{fig:ccdf}. In this plot, we can interpret the differences in Structure and Participation between Reddit and congressional hearings regarding abortion. 

\subsubsection{Robustness.}\label{sec:robustness} To evaluate the sensitivity of our Deliberation Intensity Score to methodological choices, we compare calculating the structural features using two conceptually different clustering algorithms. Participation features do not utilize clustering, so they will remain the same. The Sentence Transformers graph-based Community Detection Algorithm and density-based clustering HDBSCAN are the two methods being evaluated. The details of this algorithm can be found in Section \ref{sec:viz}. For the purpose of readability, the Community Detection Algorithm will be referred to `ST', and we evaluate significance using a t-test.

The structural features demonstrated significant differences between methods $(p < 0.05)$, with ST identifying more diverse clusters (mean difference of 0.027 to 0.267) but lower narrative distinctness (mean differences of -0.049 to -0.182). Although we find a majority of the structural features to be different, the overall Deliberation Intensity Score remained robust across the different methods and did not exhibit any statistically significant difference. This suggests our DIS metric may be capturing latent aspects of deliberation that transcend specific methodological choices. The metric serves as an exploratory tool for understanding argumentative patterns rather than an absolute measure of deliberative quality, providing comparative insights into structural differences. We acknowledge the inherent complexity of validating deliberative measures, and future work will explore alternatives in clustering or structural design approaches to further validate these findings. Possibilities for how different sentence embedding models or clustering parameters affect the metrics stability will provide valuable insights into the reliability of the Deliberation Intensity Score.

\subsection{Quantifying the Evolution of Discourse}
The problem motivating this analysis is, given a specific legislative hearing or social media discussion, do arguments build off of one another or do they diverge in different directions? How can we use computational methods to model this phenomenon? We propose a method to model the convergence or divergence of arguments over the course of a debate. 
This modeling occurs at the discourse event level (social media thread or legislative hearing). In order to analyze this sequence as a temporal event, we define the relative sequential difference between two spoken statements or written comments to be a fixed time sequence. For example, even if the next user to comment in a social media thread, after an initial comment was made, happens 10 hours later we would still consider the time step to be ``1''. Understandably, this may overlook certain interesting temporal relationships, but for the sake of this work we seek to standardize our temporal dimension and as such, we also treat legislative proceedings in the same way. This allows our methods to be generalizable for both online and offline debates of different argument lengths and debate duration.

The first step is the embedding of arguments using the `all-mpnet-v2' Sentence Transformer model discussed in section \ref{section:arg_theme}. This results in a vector embedding space for each argument in 768 dimensions. Intuitively, this transformation will capture semantic information that is vital for measuring the similarity and differences between arguments. 

From here, two potential design choices for calculating this evolution become apparent. When comparing the trajectory of arguments throughout a discussion one might use an anchor of \textit{k} initial arguments to compare later arguments to. Though this approach may be valid, it assumes that the catalyst for discussion is these first few arguments, which trivializes properties of complex debate. The assumption that discussions always start from their most coherent point does not consistently hold. The second approach is the use of an adaptive sliding window to calculate the local average semantic distance of arguments ensuring all time points are treated equally. We decided to use the adaptive sliding window, as it better represents our goal to capture thematic development and local structures of debate while remaining robust to noise or outliers. This enables us to accurately assess the coherence of deliberation at each point and reflects the \textit{dynamic} nature of deliberation. 

\textbf{Adaptive Sliding Window.} For each thread or hearing, we calculate an adaptive window size $w = \lfloor\sqrt{n}\rfloor$, where $n$ is the total number of arguments, bounded by minimum and maximum sizes to ensure meaningful analysis. For a discussion with embedding vectors $\mathbf{E} = \{\mathbf{e}_1, ..., \mathbf{e}_n\}$, we create a sequence of overlapping windows $W_i = \{\mathbf{e}_i, ..., \mathbf{e}_{i+w-1}\}$ for each position $i$ in the sequence. For each window $W_i$, we calculate its semantic center $\mathbf{c}_i = \frac{1}{w}\sum_{\mathbf{e} \in W_i} \mathbf{e}$ and compute cosine distances from all arguments to this center: $d_{ij} = 1 - \frac{\mathbf{e}_j \cdot \mathbf{c}_i}{\|\mathbf{e}_j\| \|\mathbf{c}_i\|}$. The final sequence of distances is smoothed using an exponentially weighted moving average with span $w$, producing a measure of semantic coherence over time. Figure \ref{fig:evolution} visualizes these distances for Reddit threads and congressional hearings regarding abortion, showing how argument similarity evolves throughout the discussions.

\subsection{The DALiSM platform}
\label{sec:viz}

The work culminates with the implementation of DALiSM as a platform, shown in Figure \ref{fig:delib}. 
The platform is visually informative, interactive, and easy to use. For a given deliberation platform, we can model both the dynamics of deliberation and the discourse event level quantification of the argumentation that takes place.
The platform implements the algorithmic solutions presented here, but it is designed to be modular, so that its components can be easily improved in the future.

% \miii{Reduce, shrink or remove the below or keep select choices}
For the visualization of similar argument themes, we design DALiSM with a modular approach, enabling the use of any community detection or clustering algorithm that best fits the use case. In this paper, the Sentence Transformers threshold-based community detection algorithm is applied to the set of argument embeddings present in each discourse event for visualization and clustering purposes. For each pair of embeddings, if the cosine similarity is greater than the defined threshold, a network edge connects the two embeddings. Finally, all connected components in the resulting graph are assigned to a cluster or ``community", which is a set of semantically similar arguments of any size. As defined previously, we consider these sets of semantically similar arguments to be ``narratives", similar to our definition in Section 2.4.  Each narrative cluster is then summarized using LLama 3.1-8B \cite{dubey2024llama3herdmodels} and few-shot prompting to generate a concise summary of key points being argued.

\textbf{Impact.} This web-app aids researchers in understanding the flow of changing arguments or ideas in a discourse. Using our dashboard, as shown in Figure \ref{fig:interactive}, researchers can generate an interactive plot for a selected discourse event. If the data source is a legislative hearing, for example, the chart will populate with member and witness information, which can be observed if hovering over any dialogue box. Member information includes the speaker's political party, whether their party was in the majority, the state they represent, and a distilled key point summary of the argument presented in each statement they make.\footnote{This distillation is performed with a complementary Llama 3.1-8B model tasked with performing participant level summarization.} Finally, all clusters have a concise summary contained within easy-to-read boxes below the chart, which provides researchers with an at-a-glance understanding of all the different arguments made in the selected hearing. The dashboard visually presents the arguments chronologically and color-coded by community, where each color represents a distinct common viewpoint in the set. This visualization is useful in various ways: for example, it allows analyzing who proposed an idea or made a new argument for the first time in a discourse event, which speakers and how many other speakers adopted and reiterated the argument subsequently, whether a discourse event is characterized as convergence toward the same idea or divergence of ideas as it progresses, among other possibilities.
\begin{figure*}[htbp] 
\centering
\begin{minipage}{\linewidth}
    \includegraphics[width=\textwidth]{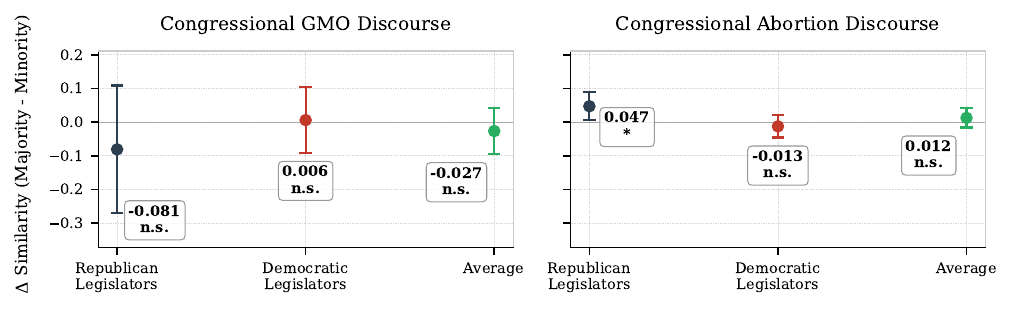}
    \Description{The interactive dashboard displays various metrics and visualizations from the U.S. Congressional Hearing titled "Revoking Your Rights: The Ongoing Crisis in Abortion Care Access." The interface includes argument clusters, discussion timelines, and participant statistics.}
    \caption{Do legislators and witnesses use similar arguments, and how is this affected by which party is in control? This figure quantifies the difference in argument-level agreement in congressional hearings when the legislator is in the ``party of power" versus when they are not. A positive value indicates greater similarity when the party is in control of the chamber, possibly implying that legislators in power call witnesses who agree with them.
    %, and a negative value indicates more similarity when the minority party is in control.
    }
    \label{fig:majsim_abortion}
    \centering
\end{minipage}
\end{figure*}

Our proposed visualization tool can be customized for any discourse text corpus where multiple speakers take turns and make arguments beyond the examples used in this study. For example, researchers can upload any text data of interest, such as floor speeches in a legislature or jury deliberations. As long as the data meets formatting requirements (e.g., each statement parsed and recorded as a row in a CSV file with speaker information labeled for each statement), they can visualize any discourse event to analyze and illustrate the flow of ideas.

\section{Case Study}

The overarching goal is to demonstrate the usefulness of DALiSM by showing how a political scientist would use  DALiSM to answering research questions of interest. To achieve this, we apply DALiSM to:
(a) debates  
in U.S. congressional committee hearings and,
(b) discussions on Reddit, in order to demonstrate that DALiSM can be generally applied to deliberative events of any kind. In addition, we focus on two controversial topics: (a) abortion,  and (b)  genetically-modified organisms (GMOs).

We begin by presenting a detailed overview of the datasets used in our case study, including their collection methods. Our case study then proceeds in three stages: (a) an introductory analysis of deliberation patterns in U.S. congressional committee hearings, (b) a comparative investigation of deliberation dynamics across platforms and topics, and (c) a demonstration of the capabilities of DALiSM through an analysis of argument cluster summaries, showcasing its potential for extracting meaningful insights. We begin by analyzing U.S. congressional committee hearings to establish a contextual baseline of insights derived from structured, formal deliberative settings. As mentioned, congressional hearings remain relatively unexplored in computational argumentation research.

\subsection{Legislative Proceedings Dataset}
We test our methods on a collection of U.S. congressional committee hearings on abortion and genetically modified organisms (GMOs). To identify hearings on each of these particular policy issues, we started with a seed keyword (e.g., abortion and genetically modified organisms or GMOs) and selected all the hearings that contained these seed words three times or more. Then, we used a BERT-based keyword extraction tool (BERTopic) as well as the RAKE Python package designed for keyword extension, and treated each statement in a hearing as a document to identify topics commonly talked about in order to expand our keyword set. Using the expanded set of keywords for each policy issue, we counted the number of appearances of these keywords in each hearing. Then, we selected hearings in which these keywords together appeared more than ten times. We asked GPT-4 whether each hearing we pre-selected was primarily about the given policy issue based on the first 3,000 words spoken in the hearing. Based on the results from these queries, we eventually identified 14 hearings on GMOs and 26 on abortion. There are 63 unique witnesses and 133 members of Congress for the GMO hearings, with a total of 2,810 statements made. For abortion hearings, there are 109 unique witnesses, and 193 members with a total of 7,363 statements made.

\subsection{Social Media Dataset}
We collected Reddit posts relevant to the two policy issues, GMO and abortion. In total, we collected 32,154 Reddit threads for the analysis. Our initial corpus contains 20,679,696 comments from 8 abortion-related Subreddits and 630,785 comments from 4 GMO-related Subreddits. Each Subreddit was selected based on its sustained activity level, self-provided community description, and relevance to the topic of interest. The data was collected from the inception of each community, as early as 2008 until February 2023. To further process the data, we selected threads whose title contained one of the keyword seeds identified previously, and had at least 50 comments within the discussion. Next, we randomly selected 500 threads to perform our analysis. For the topic of abortion, this resulted in a total of 117,907 unique arguments and for GMOs, 3,817. \raggedbottom

\subsection{U.S. Congressional Committee Findings}

\textbf{Partisan control of witness invitations and the argument similarity between members and witnesses.} Simply put, is the selection of witnesses in the proceedings targeted and biased? Congressional scholars assume that hearings are a venue where committee members invite witnesses who reflect the view of the members, especially the majority party members of the committee, as they tend to control the selection of witnesses \cite{degregorio1992, huitt1954, park2017}. To test whether this belief holds empirically, we utilize our argument similarity tool which quantifies the semantic similarity between arguments on continuous scale from 0 to 1. We then calculate the similarity of the arguments made by witnesses in their testimony and those made by members in a given hearing. This dyadic similarity was computed by members' parties, whether their party held the majority status in the chamber, and by the policy issue area we considered. The difference in average argument similarity between majority and minority status was then used to test whether party members tend to make arguments more similar to those conveyed in witness testimonies when they are in the majority and control whom to hear than when they are in the minority. Evidence supporting this pattern would suggest that majority party members tend to invite witnesses who would reinforce their predispositions on a given policy issue. We use a t-test to examine this hypothesis. 

While this pattern is absent in GMO hearings, it appears in abortion hearings. We find a statistically significant difference in the republican legislator-witness argument similarity between when Republicans controlled the majority (and thus largely managed the invitation of witnesses) and when they were in the minority. When they can select witnesses, they tend to choose those whose arguments align more closely with their own.  The results are presented in the right panel of Figure \ref{fig:majsim_abortion}. This finding suggests that in our datasets, compared with Democrats, Republicans seem to be more strategic in using hearings as an opportunity to reinforce their partisan views through witness testimonies, especially on highly partisan, salient issues, such as abortion. However, these efforts become tenuous on less prominent issues, such as GMOs.

\begin{figure*}[htbp] 
\centering
\begin{minipage}{\textwidth}
\centering
    \includegraphics[width=0.70\textwidth]{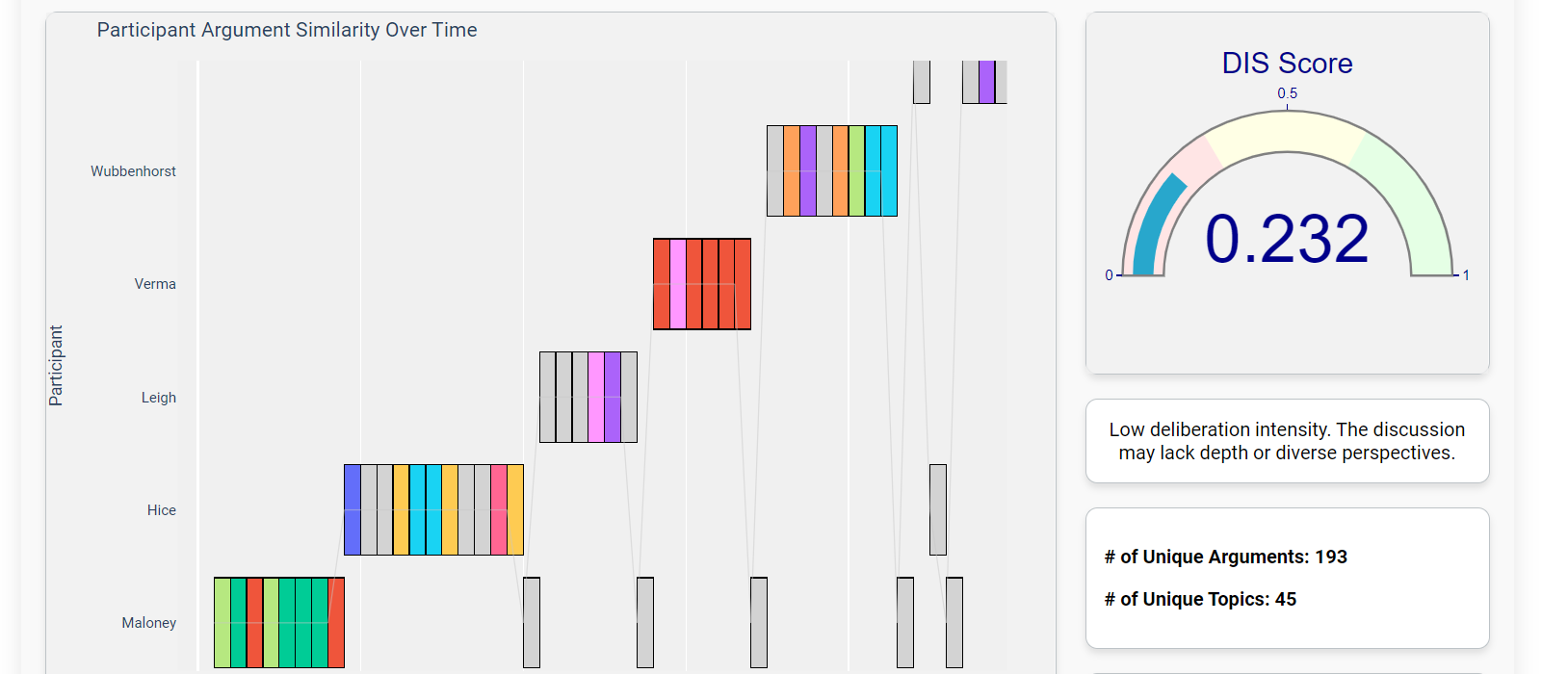}
        \Description{A screenshot of an interactive dashboard displaying analytical metrics and visualizations from the U.S. Congressional Hearing titled "Revoking Your Rights: The Ongoing Crisis in Abortion Care Access." The dashboard includes data visualizations, argument summaries, and participant metrics.}

    \caption{Interactive dashboard of the U.S. Congressional Hearing: ``Revoking Your Rights: The Ongoing Crisis in Abortion Care Access".}
    \label{fig:interactive}
\end{minipage}
\end{figure*}

\textbf{Identifying argumentative legislators. }
Interestingly, for the abortion hearings we find that Democrats exhibit the highest argumentativeness, whereas for GMO hearings, witnesses make up the population with the highest argumentative score. This suggests that legislators tend to hold hearings to present their own arguments on hot-button issues such as abortion, whereas they tend to hold hearings to seek expert information and advice from field practitioners on technical issues such as GMOs. In line with our ethical considerations statement and to further study representation in democratic societies, we explicitly mention the elected officials active in the congressional hearing. The following is a list of the five most argumentative speakers on each policy issue. In abortion hearings: Jimmy Gomez (D-CA), Richard Edmund Neal (D-MA), an unnamed bureaucrat witness, Mark James DeSaulnier (D-CA), and Shontel Brown (D-OH). In GMO hearings: Witnesses representing trade associations, corporate and non-profit organizations, and Vicky Hartzler (R-MO). \raggedbottom

\subsection{Cross-Platform Analysis}\label{sec:crossplatform} We use our systematic approach to identify differences in deliberation dynamics across (a) platforms, and (b) topics. This work does not investigate a per-hearing and per-subreddit argumentative analysis. As a proof of concept, we compare deliberation occurring in congressional hearings versus Reddit on the topics of abortion and GMOs. For Reddit discussions featuring branching or nesting comment structures typical of social media platforms, we transform the discussion into a time series by ordering comments chronologically by timestamp. This creates a streamlined, easy-to-follow argumentative flow.

\textbf{Congressional vs. Reddit Discourse Evolution.}
Using our methodology for quantifying the evolution of arguments over the course of a discourse event, we are able to visualize how a discussion unfolds across different platforms as shown in Figure \ref{fig:evolution}. We further analyze the differences between the platforms' deliberation dynamics by investigating the statistical difference in semantic discourse evolution between platforms. We independently analyze discussions regarding abortion and GMOs from Reddit and congressional hearings. Using the Kolmogorov-Smirnov test, we find that the differences between platform dynamics are statistically significant $( p < 0.05)$ for both abortion and GMO discussion, indicating a fundamental difference between platforms. Calculating the effect size\footnote{Effect sizes (Cohen's $d$) measure the standardized difference between means, where $|d| \geq 0.2$ indicates a small effect, $|d| \geq 0.5$ a medium effect, and $|d| \geq 0.8$ a large effect.}, we find that the difference in semantic evolution between platforms differs substantially for abortion, with a difference of 3.41 standard deviations, and moderately for GMOs with a difference of 0.73 standard deviations.  

Next, we  calculate the overall trend of the discourse trajectory per platform using linear regression to fit the data and find the slope. We find that Congress has a positive trend of 0.079 for abortion and 0.124 for GMOs while Reddit has a slightly negative trend of -0.002 for abortion and -0.015 for GMOs. This indicates fundamentally different evolutionary patterns, with Congress showing a steady increase in semantic distances and Reddit showing a slight regression towards lower semantic distances. However, if we were to calculate the volatility of each platform, Reddit shows considerably higher volatility with a value of 1.325 for abortion compared to 0.414 for congressional hearings. For the topic of GMOs, Reddit's volatility is almost 6 times that of Congress, with a value of 3.852 compared to 0.674 respectively. Furthermore, by dividing our normalized time into three phases and calculating the volatility for each phase, we can quantify the increase in volatility in the third phase, or towards the end of discussion for Reddit for both topics. 

\begin{figure}[t] 
% \begin{minipage}{\linewidth}
\centering
    \includegraphics[width=\linewidth]{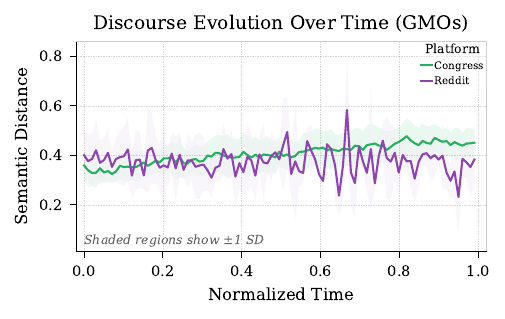}
     \Description{A line chart showing the evolution of discourse on Reddit and congressional hearings regarding GMOs, highlighting significant volatility on Reddit compared to congressional hearings.}
    \caption[Discourse Evolution for GMO Discussions]{The Evolution of Discourse on Reddit and congressional hearings regarding the topic of GMOs. We observe GMO discourse to be incredibly volatile compared to Congress, with an increase in volatility towards the end of discussions.}
    \label{fig:evolution_gmo}
% \end{minipage}
\end{figure}

These findings quantify the undeniable trend of deliberations diverging as discussion progresses, with Reddit tending to be more volatile than congressional hearings. Specifically, for discussions regarding GMOs, the landscape on Reddit is extremely diverse and divergent compared to that of congressional hearings, as seen in Figure \ref{fig:evolution_gmo}. Furthermore, while hearings trend towards higher semantic differences over the course of deliberation, these divergences are very stable. On the other hand, Reddit discussions are subject to higher moment-to-moment variation (higher volatility) with an overall pattern of stability (flat trend). Lastly, we find these differences between platforms to be statistically significant. 

\textbf{Congressional hearings foster more structured but less participatory deliberation than Reddit.} Using our Deliberation Intensity Score and its composite structural and participant properties, we find substantial differences between Reddit and congressional hearings. The following analysis demonstrates an introductory utility of our proposed metric. For abortion discussions, congressional hearings demonstrate significantly higher narrative diversity and distinctness compared to Reddit with a mean difference of -0.082 and -0.547 respectively. However, Reddit exhibits stronger participant engagement, with debater diversity and argumentativeness showing a large positive effect size of 0.862 and 1.024 respectively. This pattern suggests that while congressional hearings maintain more structured and distinct argumentative discussion, Reddit fosters broader participation and more frequent argumentation. Similar platform differences emerge in GMO discussions, with congressional hearings again showing higher narrative diversity and distinctness; however, the effect sizes for these structural properties are even larger in GMO discussions. This may indicate fundamental institutional properties that are effective at maintaining more focused deliberation. Interestingly, GMO discussions on Reddit have mixed participation differences, with Reddit having higher debater diversity but low argumentativeness compared to congressional hearings.

When examining a platform's cross-topic Deliberation Intensity, we observe abortion discussions on Reddit consistently exhibiting higher intensity across all components compared to GMO discussions, with particularly strong differences in argumentativeness, and overall DIS score. This suggests that controversial social issues may generate more engaged and structured deliberation in public forums than technical policy topics, reflected by A 25-point difference in argumentativeness. In contrast to Reddit, congressional hearings show consistency across topics, with no statistically significant differences between abortion and GMO discussions in any component (all $p>0.05$). This topic-invariance in Congressional Deliberation Intensity suggests the institutional structures and procedural norms of legislative proceedings effectively standardize deliberative patterns regardless of the subject matter. 

These findings illuminate the complementary strengths and weaknesses of different deliberative venues, with formal legislative settings excelling in consistent deliberation and informal online platforms enabling broader participation.

\subsection{DALiSM Platform Example}

\textbf{Understanding the congressional hearing: ``Revoking Your Rights: The Ongoing Crisis in Abortion Care Access.''} 
To showcase our interactive dashboard that visualizes the arguments in the order presented by each speaker during a hearing session, we randomly selected one hearing on abortion, titled ``Revoking Your Rights: The Ongoing Crisis in Abortion Care Access.'' This hearing was held by the House Committee on Judiciary on May 18, 2022, when the Democratic Party had the majority control over the chamber. Shown in Figure~\ref{fig:interactive} is the chronological argument exchange in this hearing and corresponding Deliberation Intensity Score.
To demonstrate the content generated by DALiSM, for the selected hearing the first three argument clusters are as follows:

\begin{quote}
    ``The development of abortion bans has far-reaching consequences, exacerbating existing inequities and worsening health outcomes, particularly for women of color. By restricting access to abortion care, these bans can lead to increased maternal mortality rates, worsened economic outcomes, and generational harm.''
\end{quote}

\begin{quote}
    ``The development of a nationwide abortion ban is a clear attempt by Republicans to overturn state abortion laws and impose government control over women's bodies and choices.''
\end{quote}

\begin{quote}
    ``The fundamental right to control one's own body and make decisions about reproductive healthcare is a core American value, and no one deserves to be judged for seeking an abortion."
\end{quote}

Each cluster appears as a color-coded card below the main visualization, with colors matching the hues assigned to its compositional arguments in Figure \ref{fig:interactive}. The complete cluster visualization is omitted due to space constraints.
% readers may refer to Appendix A for a comprehensive visual representation of the clustering results.

\subsection{Related Works}\label{sec:related}
\textbf{Legislative Discourse Analysis.}
Legislative discourse is well-documented in many democratic countries, offering scholars numerous opportunities to analyze its nature. In the U.S. Congress, floor speeches, congressional committee hearing transcripts, and the Congressional Record have been key sources for discourse analysis. With advances in machine-learning-assisted text analysis, researchers have developed various approaches to studying legislators' speech patterns. For instance, \cite{park2021} measures legislators' grandstanding tendency using congressional committee hearing transcripts. \cite{gentzkow2019} analyzed Congressional Record speeches to assess partisanship, while \cite{diermeier2012} explored the ideology of U.S. Senators through their speeches. In a similar vein, but in Canadian parliamentary discourse, \cite{naderi2016argumentation} investigates the ideological framing of policy issues for speakers in parliamentary discourse. These studies typically use supervised learning methods to classify or score latent traits in speeches. However, to our knowledge, none of the existing studies have applied argument-mining techniques to legislative discourse. Our paper is the first to take this approach, enabling a more in-depth analysis of legislative discourse at the argument level.

\textbf{Social Media Discourse Analysis.}
Since its inception, social media discourse has been consistently studied across a wide variety of platforms. Its ability to connect, divide, empower, and radicalize is an important area of study and consideration. For example, \cite{chang2019trouble} investigates discourse as a derailment into toxic exchanges between participants, seeking to quantify and model the discourse in social media data. When it comes to modeling social media behavior, \cite{wei2019modeling} proposes a model for analyzing the structure of social media conversations over time for the task of rumor and veracity prediction. Discourse dynamics and the interactions therein were analyzed by \cite{mukherjee2012analysis}, which presents a framework for modeling agreement and disagreement in online debates, and the psycholinguistic accommodation during the debates. Further work by \cite{strnadova2013characterizing} provides a network-based analysis for quantifying and categorizing discussion features and dynamics. However, to the best of our knowledge, this paper is the first to take computational argumentation methods and apply deliberation modeling as a generalizable framework.  

\section{Discussion}
In this section we discuss the impact, limitations, and extensions of our work. 

{\bf a. DALiSM as a research tool for social sciences.}
We envision our platform to be used as a tool for social scientists to easily analyze any set of legislative or social media discourse that is readily available. We designed DALiSM to be as user-friendly as possible. First, it is an online platform, thus not requiring any installation or computational requirements. Second, we enable users to upload data and obtain initial results without having to specify any obscure parameters. At the same time, our interactive operation, with more functionality forthcoming, enables researchers to explore specific phenomena and areas.

{\bf  b. Can it work with other types of data?}
Our methods were designed to be generalizable, applicable to any kind of dialogical data source. Here we discuss two cases where the data may not come with information that so far we assumed is available. First, contributions to a discourse event have to be organized in a sequentially, meaning each contribution must be assigned a timestamp or an explicit order of appearance. If the temporal ordering does not exist or is not known, the tool can still work with the ordering that the researcher imposes on their data. Second, if the author information is absent or anonymized, our tool can attribute each contribution to a new author each time. We have shown that DALiSM works with formal legislative hearings and informal discussion forums. The current form of dialogical representation, visually, is chronological, which uses the timestamp to order the appearance of arguments and non-arguments. For now, and for the scope of this work, the branching of dialog that occurs in some online discussions is not considered. These `side conversations' introduce complex considerations and implications that we leave for future research. 

{\bf c. Are the observations generalizable?}
The main goal of this work is to present our approach and publicize its implementation. The results that we report here are first and foremost an indication of the types of outcomes that can be obtained with the use of DALiSM. At the same time, we were meticulous in collecting and analyzing our results, which makes us confident in their accuracy for the datasets we analyzed.

{\bf  d. Limitations in NLP capabilities.}
Despite recent advancements in Natural Language Processing (NLP) methods, the inherent complexity, vagueness, and context-dependent nature of human language and communication pose significant challenges for computational analysis. Our deliberation analysis platform, DALiSM, while pushing the boundaries of current state-of-the-art methods in this field, is constrained by these fundamental limitations. In particular, DALiSM is inherently limited by the capabilities of the computational tools employed. In the context of social media analysis, the framework is unable to process and derive meaning from visual media, such as images or videos, which often play a crucial role in online argumentative discourse. Moreover, DALiSM may struggle to accurately interpret modern linguistic phenomena, including slang, emojis, or memetic terminology, which human observers can more readily comprehend.

The modular design of our approach allows for continual improvement and integration of the latest NLP advancements and research findings. However, it is crucial to acknowledge that many state-of-the-art NLP tools and techniques often exist independently, developed within research silos or tailored for specific domains. This fragmentation poses significant challenges for creating a unified, comprehensive framework capable of robustly handling the full spectrum of human deliberation dynamics. To address these limitations and advance the field of computational social science, bridging the gaps between disparate methodologies and fostering collaboration across research communities will be essential. By leveraging the strengths of various approaches and adapting them to the unique challenges of multi-platform discourse analysis, future research can contribute to the development of more sophisticated and integrated natural language processing techniques that effectively capture the nuances of human communication modalities and linguistic innovation prevalent in online deliberative environments.

{\bf e. Ethical Considerations.} We follow best practices according to the ACM code of ethics and professional conduct.
First, we only analyze public data. Both hearings and online forums are provided for public access, and the participants are aware that their opinions are public. 
Second, we mostly report aggregate results, focusing on group behavior rather than individual actions.
We strictly adhere to consent-based data usage, ensuring no unauthorized personal information is revealed or used within our analysis. However, we are transparent about the arguments of legislators. We feel that by virtue of their role as elected officials, they should be held accountable to a higher standard of transparency as is the basis of an ethical system. 
Third, our ability to summarize group behaviors and tendencies could provide insights on deliberation dynamics.
In the data-mining field,
any capability and knowledge can be used or misused.
However, we firmly believe that our work can provide transparency and knowledge that can help 
establish safeguards to uphold effective and democratic discourse and processes. Our goal is to strengthen the resilience of \textit{online} and \textit{offline} discourse environments against unhealthy polarization and disinformation campaigns.

\section{Conclusion}

This paper introduces DALiSM, our approach that captures the dynamics of deliberation in diverse and multi-party spaces, such as legislative hearings and social media discussion. A key contribution is our novel Deliberation Intensity Score, which quantifies argumentative diversity and participant engagement while acknowledging the inherent complexity of measuring deliberation. This metric, along with our proposed visualization strategies, enables systematic comparative analysis of deliberative dynamics. Our discourse analysis framework can be applied to future research on the efficacy of legislative deliberation systems, legislators' behavior, the influence of external groups on lawmaking processes, and even linguistic queries. We hope that these illustrations can stimulate research on other measures and metrics of discourse that might be of interest to the fields of political science, public policy, sociology, psychology, and linguistics.

% \section{Acknowledgments}

\begin{acks}
    This work was supported by CDFA grant 2021 Specialty Crop Block Grant Program H.R. 133 No. 000705282.

\end{acks}
\balance
% \printbibliography{custom}
\bibliographystyle{ACM-Reference-Format}
\bibliography{main}

\end{document}